\documentclass[aps,prb,twocolumn,superscriptaddress,showpacs,floatfix,amsmath,amssymb]{revtex4}

\usepackage{graphicx}
\usepackage{dcolumn}
\usepackage{bm}
\usepackage[below]{placeins}

\begin{document}

\title{Interplay between Fe and Nd magnetism in NdFeAsO single crystals}

\author{W.~Tian}
\affiliation{Ames Laboratory, Ames, Iowa 50011, USA}

\author{W.~Ratcliff II}
\affiliation{NIST Center for Neutron Research, National Institute
of Standards and Technology, Gaithersburg, Maryland 20899, USA}

\author{M.\,G.~Kim}
\affiliation{Ames Laboratory, Ames, Iowa 50011, USA}
\affiliation{Department of Physics and Astronomy, Iowa State
University, Ames, Iowa 50011, USA}

\author{J.-Q.~Yan}
\affiliation{Ames Laboratory, Ames, Iowa 50011, USA}

\author{P.\,A.~Kienzle}
\affiliation{NIST Center for Neutron Research, National Institute
of Standards and Technology, Gaithersburg, Maryland 20899, USA}

\author{Q.~Huang}
\affiliation{NIST Center for Neutron Research, National Institute
of Standards and Technology, Gaithersburg, Maryland 20899, USA}

\author{B.~Jensen}
\affiliation{Ames Laboratory, Ames, Iowa 50011, USA}

\author{K.\,W.~Dennis}
\affiliation{Ames Laboratory, Ames, Iowa 50011, USA}

\author{R.\,W.~McCallum}
\affiliation{Ames Laboratory, Ames, Iowa 50011, USA}
\affiliation{Department of Materials Science and Engineering, Iowa
State University, Ames, Iowa 50011, USA}

\author{T.\,A.~Lograsso}
\affiliation{Ames Laboratory, Ames, Iowa 50011, USA}

\author{R.\,J.~McQueeney}
\author{A.\,I.~Goldman}
\affiliation{Ames Laboratory, Ames, Iowa 50011, USA}
\affiliation{Department of Physics and Astronomy, Iowa State
University, Ames, Iowa 50011, USA}

\author{J.\,W.~Lynn}
\affiliation{NIST Center for Neutron Research, National Institute
of Standards and Technology, Gaithersburg, Maryland 20899, USA}

\author{A.~Kreyssig}
\affiliation{Ames Laboratory, Ames, Iowa 50011, USA}
\affiliation{Department of Physics and Astronomy, Iowa State
University, Ames, Iowa 50011, USA}

\date{\today}

\begin{abstract}

The structural and magnetic phase transitions have been studied 
on NdFeAsO single crystals by neutron and x-ray diffraction 
complemented by resistivity and specific heat measurements.  Two 
low-temperature phase transitions have been observed in addition 
to the tetragonal-to-orthorhombic transition at $T_S\sim$\,142\,K 
and the onset of antiferromagnetic (AFM) Fe order below 
$T_N\sim$\,137\,K.  The Fe moments order AFM in the well-known 
stripe-like structure in the (\textbf{ab}) plane, but change from 
AFM to ferromagnetic (FM) arrangement along the \textbf{c} 
direction below $T^*\sim$\,15\,K accompanied by the onset of Nd 
AFM order below $T_{\textrm{Nd}}\sim$\,6\,K with this same AFM 
configuration.  The iron magnetic order-order transition in 
NdFeAsO accentuates the Nd-Fe interaction and the delicate 
balance of \textbf{c}-axis exchange couplings that results in AFM 
in LaFeAsO and FM in CeFeAsO and PrFeAsO.

\end{abstract}


\maketitle

Since the discovery of high-temperature superconductivity in the 
fluorine-doped compound LaFeAsO$_{1-x}$F$_x$ with a 
superconducting transition temperature of 
$T_c$\,=\,26\,K\cite{Kamihara-2008}, enormous efforts have been 
made to understand the crystallographic, magnetic, and 
superconducting properties of the $R$FeAsO compounds 
($R$\,=\,rare earth elements) and their doped superconducting 
derivatives with $T_c$s up to 
$\sim$\,55\,K\cite{Cruz-2008,Sefat-2008,Ren-Pr1111-2008,Ren-Nd1111-2008,Chen-2008,Ren-Euro-2008,Yang-Supsci-2008,Bos-2008,Liu-PRL-2008,Wang-Euro-2008,Chen-PRB-2008,Qiu-PRL-2008,Huang-2008,Marcinkova-2009,Ryan-2009,Marcinkova-2010,Malavasi-2010}. 
The parent $R$FeAsO compounds undergo a 
tetragonal-to-orthorhombic phase transition and show 
antiferromagnetic (AFM) order of Fe and $R$ 
moments\cite{Cruz-2008,Chen-PRB-2008,Qiu-PRL-2008,Huang-2008,Marcinkova-2009,Ryan-2009,Marcinkova-2010,Malavasi-2010}. 
Although doping suppresses both the lattice distortion and Fe 
magnetic order, magnetism is believed to play a major role in the 
superconducting pairing mechanism in pnictide superconductors 
demonstrated, e.\,g., by the observation of a magnetic resonance 
in the superconducting state\cite{Christianson-nature-2008}. 
Detailed know\-ledge of the magnetic order and crystallographic 
structure is key to gaining insight into the magnetic and 
magneto-elastic coupling, but has been hampered, so far, by the 
absence of sizeable $R$FeAsO single crystals.

Here we report on a neutron and x-ray diffraction study of 
recently available\cite{Yan-APL-2009} NdFeAsO single crystals 
complemented by resistivity and specific heat measurements.  In 
prior neutron powder studies, the tetragonal-to-orthorhombic 
lattice distortion has been observed below 
$T_S\sim$\,150\,K\cite{Qiu-PRL-2008} followed by the onset of the 
well-known stripe-like AFM order of Fe below 
$T_N\sim$\,141\,K\cite{Chen-PRB-2008}.  The onset of the Nd AFM 
order was reported for temperatures below 
$T_{\textrm{Nd}}\sim$\,2\,K in neutron powder diffraction 
experiments\cite{Qiu-PRL-2008,Marcinkova-2009,Malavasi-2010}. 
However, resistivity \cite{Cheng-PRB-2008} and muon-spin 
relaxation \cite{Aczel-PRB-2008} studies indicate a 
low-temperature phase transition at 5-6\,K.  Our NdFeAsO single 
crystals show two low-temperature transitions in addition to the 
high-temperature transitions at $T_S\sim$\,142\,K and 
$T_N\sim$\,137\,K.  Below $T^*\sim$\,15\,K, the Fe moments 
undergo a strongly hysteretic magnetic transition from AFM 
arrangement to a ferromagnetic (FM) arrangement along the 
\textbf{c} direction but unchanged AFM stripe-like order in the 
(\textbf{ab}) plane.  The Nd moments order AFM below 
$T_{\textrm{Nd}}\sim$\,6\,K, coupled to the Fe magnetic order by 
a common magnetic unit cell, indicating a complex interplay 
between Nd and Fe magnetism.

NdFeAsO single crystals were grown out of NaAs flux as described 
in Ref.\,\onlinecite{Yan-APL-2009}.  Batches with mm-sized single 
crystals and masses up to 7\,mg were carefully examined by 
wavelength dispersive spectroscopy (WDS) in a JEOL JXA-8200 
Superprobe electron probe microanalyzer and x-ray powder 
diffraction.  As in Ref.\,\onlinecite{Yan-APL-2009}, WDS 
measurements on different samples from the same batch show no 
noticeable compositional variation and confirmed, within the 
limits of the measurement, the atomic ratio of 1:1:1:1.  No 
impurity phases were detected beyond the NaAs flux.  

Specific heat and resistivity were measured in a Quantum Design 
Physical Property Measurement System \cite{NIST} by the 
relaxation method and the four-probe technique using Epotek H20E 
silver epoxy contacts on platinum leads.  The homogeneity of the 
batches was examined by resistivity measurements of several 
single crystals showing similar behavior.  The reproducibility 
and excellent homogeneity of the grown material were additionally 
confirmed by neutron diffraction measurements on polycrystalline 
samples consisting of more than 300 crystals.  They showed 
well-defined transition temperatures and Bragg peaks in excellent 
agreement with the single crystal measurements reported here.  To 
check potential ferromagnetism, the magnetization, $M$, was 
measured with increasing temperature from 5 to 20\,K for eleven 
fields, $H$, between 0.1 and 5\,T and the differential 
susceptibility was calculated at each temperature.  Parallel to 
the \textbf{c} axis, the zero-field intercept of the $M$ vs $H$ 
line characterizing a FM component was essentially temperature 
independent giving an upper limit of 0.012\,$\mu_B$/formula 
unit.  Perpendicular to the \textbf{c} axis the value was an 
order of magnitude smaller.

The lattice distortion was characterized by high-resolution x-ray
diffraction measurements using a four-circle diffractometer and
Cu-$K_{\alpha1}$ radiation from a rotating anode x-ray source,
selected by a Ge(1\,1\,1) monochromator.  The plate-like single
crystal was attached to a flat copper sample holder on the cold
finger of a closed-cycle displex refrigerator.  A mosaicity of
$\sim$\,0.03\,deg was determined from full width at half maximum
of the rocking curve of the (0\,0\,6) Bragg peak.  The
orthorhombic distortion, $\delta$, was determined as described in
Ref.\,\onlinecite{Nandi-PRL-2010} using ($\xi\,\xi$\,0) scans
through the tetragonal (1\,1\,7) Bragg peak which splits into the
orthorhombic (2\,0\,7) and (0\,2\,7) Bragg peaks below $T_S$.

The neutron diffraction experiments were carried out using the 
BT9 thermal neutron triple-axis spectrometer at NCNR, NIST.  The 
crystal with a mass of 3\,mg was mounted on a thin aluminum post, 
oriented in the orthorhombic ($h\,0\,l$) scattering plane, and 
placed in an ILL cryostat.  Rocking scans were resolution 
limited, indicating the high quality of the sample. Neutrons with 
a fixed incident energy of 14.7\,meV and a collimation of 
40'-48'-sample-40'-open were used.  Additional measurements were 
taken on a number of crystals heat treated in various ways to 
check the results, and the observed transitions were all 
consistent within uncertainties. Descriptions are given in terms 
of the orthorhombic crystallographic (nuclear) unit cell present 
below $T_S$.

\begin{figure}[!ht]
\centering\includegraphics[width=0.7\linewidth]{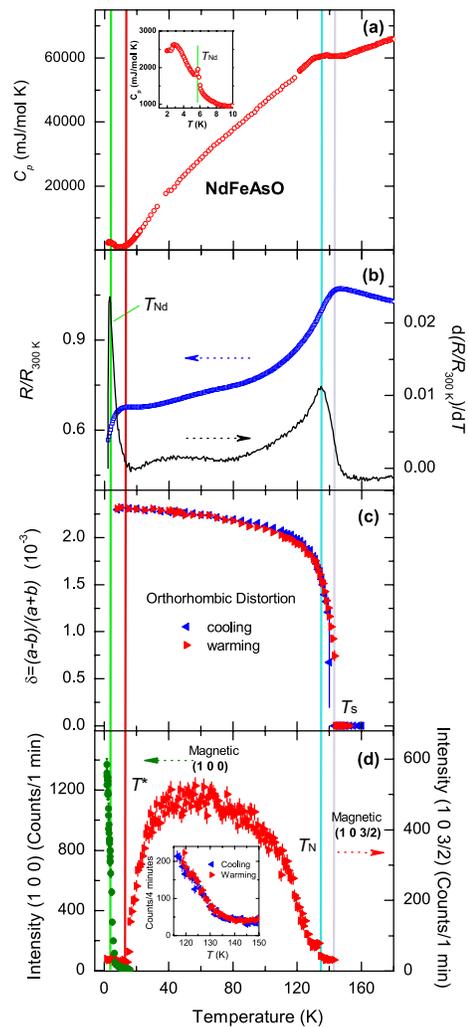}
\caption{\label{fig:microstudy}(Color online) (a) Temperature
dependence of the specific heat.  (b) Resistivity (open symbol,
left axis) and its derivative (line, right axis).  (c)
Orthorhombic distortion, $\delta$, determined by x-ray 
diffraction as described in the text.  (d) Neutron peak intensity 
for the magnetic Bragg peaks at (1\,0\,0) (left axis) and at 
(1~0~3/2) (right axis).  The insets show expanded scales.}
\end{figure}

\begin{figure}[!ht]
\centering\includegraphics[width=0.68\linewidth]{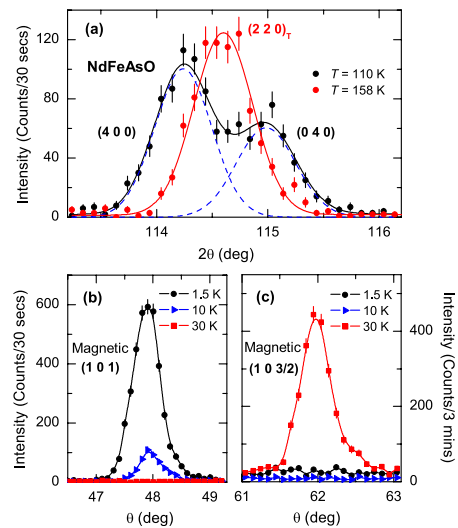} 
\caption{\label{fig:rockscan}(Color online) (a) Neutron 
diffraction $\theta$-2$\theta$ scans through the tetragonal 
(2\,2\,0) nuclear Bragg peak at $T=$\,158\,K and its splitting 
into two orthorhombic Bragg peaks at $T=$\,110\,K measured with 
reduced collimation of 10'-10'-sample-10'-10'.  Neutron 
diffraction rocking scans through the magnetic Bragg peaks at (b) 
(1\,0\,1) and (c) (1~0~3/2) at selected temperatures.}
\end{figure}

In Fig.~\ref{fig:microstudy}, we compare the four measurements as
a function of temperature.  The anomaly at $T_S\sim$\,142\,K in
specific heat and resistivity measurements correlates with the
phase transition from tetragonal $P\,4/n\,m\,m$ to orthorhombic
$C\,m\,m\,a$.  The orthorhombic lattice distortion, $\delta$,
increases below $T_S$ for decreasing temperature without further
anomalies.  Figure\,\ref{fig:rockscan}(a) shows the splitting of
the tetragonal (2\,2\,0) Bragg peak into the orthorhombic
(4\,0\,0) and (0\,4\,0) Bragg peaks from our high-resolution
neutron diffraction measurements and demonstrates the bulk nature
of the lattice distortion and the homogeneity of the sample,
through resolution-limited peak shapes and a uniform transition
temperature.

The second anomaly found in specific heat and resistivity data at
$T_N\sim$\,137\,K corresponds to the onset of the well-known AFM
stripe-like Fe order\cite{Chen-PRB-2008}. The intensity of the
magnetic (1~0~3/2) Bragg peak increases smoothly below $T_N$ [see
Fig.\,\ref{fig:microstudy}(d)].  The inset illustrates that this
transition shows no hysteresis upon cooling and warming through
$T_N$ consistent with a second order transition.

At low temperature, a weak and broad anomaly in the resistivity
can be associated with the disappearance of the magnetic 
(1~0~3/2) Bragg peak at $T^*\sim$\,15\,K signaling a change in 
the Fe magnetic order.  This behavior is demonstrated in 
Fig.\,\ref{fig:rockscan}(c) by rocking scans through the 
(1~0~3/2) peak at selected temperatures.  At the same 
temperature, $T^*$, new magnetic Bragg peaks occur as illustrated 
in Fig.\,\ref{fig:rockscan}(b) by the (1\,0\,1) peak and in 
Fig.\,\ref{fig:microstudy}(d) by the (1\,0\,0) peak.  The 
intensity of these peaks increases with decreasing temperature as 
the Fe order changes, while below $T_{\textrm{Nd}}\sim$\,6\,K the 
increase is much more pronounced.  At $T_{\textrm{Nd}}$, a sharp 
peak is also observed in the specific heat [see inset of 
Fig.\,\ref{fig:microstudy}(a)] and the temperature derivative of 
the resistivity indicating the onset of the Nd magnetic order.

\begin{figure}
\centering\includegraphics[width=0.7\linewidth]{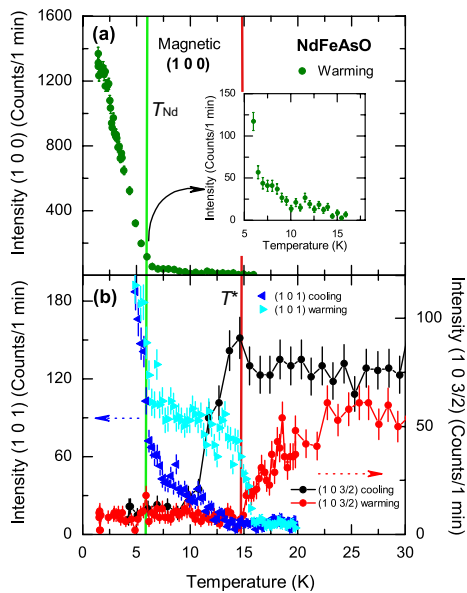}
\caption{\label{fig:magphases}(Color online) Neutron peak
intensity for the magnetic Bragg peaks at (a) (1\,0\,0) and (b) 
at (1\,0\,1) (left axis) and (1~0~3/2) (right axis).  The inset 
in (a) shows an expanded scale between $T_{\textrm{Nd}}$ and 
$T^*$.}
\end{figure}

Details of the temperature dependence for these low-temperature
transitions are shown in Fig.\,\ref{fig:magphases}.  The
transition at $T^*\sim$\,15\,K shows hysteretic behavior
illustrated in Fig.\,\ref{fig:magphases}(b), through the
temperature dependence of the magnetic (1~0~3/2) and (1\,0\,1)
Bragg peaks.  The strong hysteresis of approx.~3\,K points to the 
first order character of this transition at $T^*$. As illustrated 
in Fig.\,\ref{fig:magphases}, the (1\,0\,0) and (1\,0\,1) Bragg 
peaks follow the same trend with (i) weak intensity between $T^*$ 
and $T_{\textrm{Nd}}$, and (ii) a strong but smooth and monotonic 
intensity increase below $T_{\textrm{Nd}}$ without reaching 
saturation down to the lowest measured temperature of 1.5\,K.  
Both low-temperature transitions are reminiscent of 
Nd$_2$CuO$_4$, where the Nd-Cu coupling leads to abrupt changes 
in the Cu spin structure along with a significant induced Nd 
moment, before the Nd spontaneously 
orders\cite{Skanthakumar-1989,Lynn-1990,Sachidanandam-1997}. It 
will be interesting to determine theoretically if a similar type 
of Fe-Nd coupling is the origin of this behavior in NdFeAsO.

For the determination of the magnetic structures, a series of 5 
crystallographic and 7, 5, and 15 magnetic Bragg peak intensities 
were measured by rocking scans at $T=$\,30\,K, 10\,K, and 1.5\,K, 
respectively, and corrected for instrumental resolution.  The 
crystallographic Bragg peaks were fit using the published crystal 
structure\cite{Qiu-PRL-2008} to determine the scale factor as the 
only refinable parameter.  The value of the Fe magnetic moment 
was then determined from the intensity of the magnetic Bragg 
peaks considering equal domain averaging and using this scale 
factor and the magnetic structure models given below.  

At $T^*<T=$\,30\,K\,$<T_N$ we find the same stripe-like AFM 
structure previously reported\cite{Chen-PRB-2008}.  The Fe 
moments are collinearly aligned along the \textbf{a} axis with 
AFM arrangement in the \textbf{a} and \textbf{c} directions and 
FM in the \textbf{b} direction.  The magnetic unit cell for this 
magnetic structure is doubled along the \textbf{c} direction with 
respect to the orthorhombic crystallographic (nuclear) unit cell 
and yields the observed magnetic Bragg peaks at positions indexed 
by a propagation vector of (1~0~1/2).  The determined value of 
0.54(3)\,$\mu_B$ for the Fe moment is larger than the previously 
reported value of 0.25(7)\,$\mu_B$\cite{Chen-PRB-2008}.  The 
reason for this difference is unknown and points to differences 
between polycrystalline and single crystal samples (potential 
small deviations in stoichiometry or strain).  However, we note 
that the magnetic peaks associated with iron ordering were not 
reported in several other neutron powder scattering 
experiments\cite{Qiu-PRL-2008,Marcinkova-2009,Malavasi-2010}. 

The magnetic diffraction pattern at 
$T_{\textrm{Nd}}<T=$\,10\,K\,$<T^*$ can be indexed using a 
magnetic unit cell identical to the orthorhombic crystallographic 
(nuclear) unit cell.  In particular, no doubling along the 
\textbf{c} direction is now required, signaling a change from AFM 
ordering in this direction above $T^*$ to FM ordering below 
$T^*$. Together with the unchanged stripe-like order in the 
(\textbf{ab}) plane, this magnetic structure is simply described 
by a propagation vector of (1\,0\,0).  The intensity of all 
measured magnetic Bragg peaks can still be fit by Fe moments 
alone with a reduced moment value of 0.32(3)\,$\mu_B$ collinearly 
aligned along the orthorhombic \textbf{a} axis. 

At $T<T_{\textrm{Nd}}$, the strongly enhanced intensity of the 
magnetic Bragg peaks indicates the onset of the Nd magnetic 
order.  The description of the magnetic diffraction pattern 
requires both Fe and Nd moments to be ordered.  The positions of 
the magnetic Bragg peaks are consistent with the reported 
low-temperature magnetic 
structure\cite{Qiu-PRL-2008,Marcinkova-2009}.  However, further 
measurements are necessary for a detailed analysis of the 
magnetic structure with its extensive number of parameters.  The 
potential overlapping of Bragg peaks from twin-domains with 
different orientations together with the small orthorhombicity 
require experiments with higher resolution or the usage of 
detwinned samples. 

Summarizing, below $T_N\sim$\,137\,K, the Fe moments order AFM 
with the ubiquitous stripe-like arrangement in the orthorhombic 
(\textbf{ab}) plane.  For $T<T^*\sim$\,15\,K\, the \textbf{c} 
axis configuration changes from AFM as observed in 
LaFeAsO\cite{Cruz-2008} to FM as found in 
CeFeAsO\cite{Zhao-natmat-2008} and 
PrFeAsO\cite{Zhao-PRB-2008,Kimber-PRB-2008}.  This temperature 
dependent change in the Fe magnetic order in NdFeAsO is unique in 
the $R$FeAsO series and demonstrates that the magnetic coupling 
in the \textbf{c} direction is at the border between AFM and FM. 
Further studies of $R$FeAsO compounds with other $R$ elements are 
necessary to determine whether this delicate balance in the 
magnetic coupling is controlled by steric effects or the 
magneto-crystalline anisotropy of the $R$ element caused by 
crystal-electric field effects transferred through a complex 
coupling of the Nd and Fe magnetism.  The proximity between the 
characteristic temperatures for the order-order transition of the 
Fe magnetic order at $T^*$ to the onset of the Nd magnetic order 
at $T_{\textrm{Nd}}\sim$\,6\,K together with a common magnetic 
unit cell at low temperatures suggests the second case as the 
more likely scenario. 

Research at Ames Laboratory and Oak Ridge National Laboratory was
supported by the U.\,S. Department of Energy, Office of Basic
Energy Science, Division of Materials Sciences and Engineering
and Scientific User Facilities Division, respectively.  Ames
Laboratory is operated for the U.\,S. Department of Energy by
Iowa State University under Contract No.~DE-AC02-07CH11358.


\begin{thebibliography}{99}
\bibitem{NIST} Any mention of commercial products within this paper is for information only; it does not imply recommendation or endorsement by NIST.
\end{thebibliography}
\end{document}